\documentclass[twocolumn,showpacs,preprintnumbers,pra]{revtex4-1}
\usepackage{graphicx}
\usepackage{dcolumn}
\usepackage{bm}
\usepackage{color}

\begin{document}

\title{Semiclassical dynamics and transport of the Dirac spin}
\author{Chih-Piao Chuu$^1$}
\author{Ming-Che Chang$^1,^2$}
\author{Qian Niu$^1$}
\affiliation{$^1$Department of Physics, University of Texas at Austin,
Austin, TX 78712, USA\\
$^2$Department of Physics, National Taiwan
Normal University, Taipei, Taiwan 11677 }

\begin{abstract}
Semiclassical theory of spin dynamics and transport is formulated
using the Dirac electron model. This is done by constructing a
wavepacket from the positive-energy electron band, and studying
its structure and center of mass motion. The wavepacket has a
minimal size equal to the Compton wavelength, and has
self-rotation about the average spin angular momentum, which gives
rise to the spin magnetic moment. Geometric gauge structure in the
center of mass motion provides a natural explanation of the
spin-orbit coupling and various Yafet terms. Applications to the
spin-Hall and spin-Nernst effects are discussed.
\end{abstract}

\maketitle

\section{Introduction}

The electron has been playing a central role in modern science and
technology. It has both a fundamental charge and spin. With the
rise of spintronics, the spin degree of freedom comes to the
fore as it is beginning to be employed for data processing
as well as storage.\cite{spintronics} Much has been learned on how
to control the spin by electrical, optical as well as magnetic
means. Recently, spin transport driven by thermal gradient has
also been demonstrated \cite{spincalo,xiaoxiao}.

In this paper, we present a semiclassical theory of spin dynamics
and transport, in order to provide an intuitive picture and
effective calculation tool for such phenomena. We will focus on
the Dirac model, not only because it is fundamental to the
electron, but also it arises as effective theory of solid-state
systems such as graphene sheet\cite{graphene} and surface of
topological insulators\cite{topo}.  Therefore, this paper can
serve a dual purpose: (1) to reveal the fundamental nature of the
electron spin, and (2) to provide a simple setting for
understanding spin related dynamics and transport phenomena in
solid state systems.

The semiclassical theory is obtained by constructing a wavepacket
in the positive energy electron band following the general
framework of Culcer and Niu\cite{culcer}. We find that the
wavepacket has a minimal size equal to the Compton wavelength, and
has self rotation about its average spin, much as people imagined
when the spin was discovered\cite{uhlenbeck,lorentz}. The
self-rotation also gives rise to the spin magnetic moment showing
the fundamental orbital nature of the latter.  The center of mass
motion has a non-abealian geometric gauge structure, which is
shown to be responsible for the spin-orbit coupling as well as
various Yafet terms.  This yields a spin-dependent anomalous
velocity under an electric field, leading to the spin Hall effect.
It also yields a spin-dependent orbital magnetization that
underlies the spin Nernst effect, the spin dependent anomalous
Nernst effect.

The paper is organized as follows. First we construct the
wavepacket and analyze its structure and current profile. In
Sec.~III, we discuss the magnetic moment generated by charge
circulation within the wavepacket, and study its coupling with a
weak magnetic field. In Sec.~IV, we derive the dynamics of the
center of mass, and discuss the relation between spin-orbit
coupling and geometric gauge structure. Finally, we discuss the
spin Hall effect and spin Nernst effect in Sec.~V based on the
non-Abelian Berry curvature calculated from the Dirac theory.

\section{Dirac electron wavepacket}

When the electron spin was first discovered from the evidence of
doublets in atomic spectra, Uhlenbeck and Goudsmit\cite{uhlenbeck}
thought it as coming from the self-rotation of the electron charge
sphere. However, the idea was criticized by Lorentz\cite{lorentz},
who argued that the surface of the sphere would have to rotate
with a tangential speed at 137 times the speed of light to produce
the accurate spin angular momentum. Ever since, we were left with
no choice but to accept the spin as an abstract concept.

In 1928, Dirac formulated the Schr\"{o}dinger equation for a
relativistic electron.\cite{dirac} The Dirac equation states
\begin{eqnarray}
(-i\hbar
c\mbox{\boldmath$\alpha$}\cdot\mbox{\boldmath$\nabla$}+\beta
mc^2)\Psi({\bf r},t)=i\hbar\frac{\partial}{\partial t }\Psi({\bf
r},t),
\end{eqnarray}
where ${\mbox{\boldmath$\alpha$}\equiv\left(\begin{array}{cc}
 0 & \mbox{\boldmath$\sigma$} \\
  \mbox{\boldmath$\sigma$} & 0
\end{array}\right)}$ and $\beta= \left(\begin{array}{cc}
 \textrm{I} & 0 \\
  0 & -\textrm{I}
\end{array}\right)$ are $4\times 4$ matrices defined
by $2\times 2$ Pauli matrices $\mbox{\boldmath$\sigma$}$ and
identity matrix.

The eigenenergy states are 4-component plane waves, with a
two-fold degenerate positive energy branch,
\begin{equation}
E(q)=mc^2\sqrt{1+\frac{\hbar^2q^2}{m^2c^2}}\equiv\epsilon(q)mc^2,\label{dis}
\end{equation}
with $\hbar q=\gamma mv$ being the relativistic momentum and
$\epsilon(q)=\gamma(v)=(1-v^2/c^2)^{-1/2}$. There is also a
two-fold degenerate branch of negative eigenenergy $-E(q)$. Dirac
assumed that these states are filled to form the vacuum. A hole in
this negative energy branch is identified as a positron, the
antiparticle of the electron.

The 4-component plane-wave eigenstates are called Dirac spinors.
They can be chosen as an orthonormal set. The two spinors for the
positive energy branch are given by
\begin{eqnarray}
|u_{1}({\bf
q})\rangle=\sqrt{\frac{\epsilon+1}{2\epsilon}}\left[\begin{array}{c}
1 \\
0 \\
\frac{\hbar q_z}{mc(\epsilon+1)} \\
\frac{\lambda_c q_+}{\epsilon+1}
\end{array}\right],\cr
|u_{2}({\bf
q})\rangle=\sqrt{\frac{\epsilon+1}{2\epsilon}}\left[\begin{array}{c}
0 \\
1 \\
\frac{\hbar q_-}{mc(\epsilon+1)} \\
\frac{-\hbar q_z}{mc(\epsilon+1)}
\end{array}\right],\label{eig}
\end{eqnarray}
with $q_{\pm}=q_x\pm iq_y$. At ${\bf q}=0$, they correspond to the
two spin eigenstates with $\sigma_z=\pm 1$.

On the other hand, the two spinors for the negative energy branch
are given by
\begin{eqnarray}
|u_{3}({\bf
q})\rangle=\sqrt{\frac{\epsilon+1}{2\epsilon}}\left[\begin{array}{c}
\frac{-\hbar q_z}{mc(\epsilon+1)} \\
\frac{-\hbar q_+}{mc(\epsilon+1)} \\
1 \\
0
\end{array}\right],\cr
|u_{4}({\bf
q})\rangle=\sqrt{\frac{\epsilon+1}{2\epsilon}}\left[\begin{array}{c}
\frac{-\hbar q_-}{mc(\epsilon+1)} \\
\frac{\hbar q_z}{mc(\epsilon+1)} \\
0 \\
1
\end{array}\right].
\end{eqnarray}
In order to have intuitive picture of spin other than abstract
operator in Dirac wave equation, we study its semiclassical
dynamics by regarding a relativistic electron as a wavepacket,
which contains only the positive energy eigenstates of the Dirac
equation,
\begin{equation}
|w\rangle=\int d {\bf q} a({\bf q},t)e^{i{\bf q}\cdot{\bf r}}
[\eta_1({\bf q},t)|u_1({\bf q})\rangle+\eta_2({\bf q},t)|u_2({\bf
 q})\rangle],\label{packet}
\end{equation}
where $a({\bf q},t)=|a|e^{-i\gamma({\bf q})}$ describes the
distribution of the wavepacket in momentum space. The wavepacket
is sharply peaked at the charge center ${\bf q}_c$, and is allowed
to have an overall phase $\gamma({\bf q})$. The probability
amplitudes $\eta_1$ and $\eta_2$ describe the composition of the
wavepacket in terms of two degenerate positive energy states with
spin up and spin down. The normalization condition of the
wavepacket $\langle w|w\rangle=1$ is satisfied if $\int d {\bf q}
|a({\bf q},t)|^2=1, |\eta_1|^2+|\eta_2|^2=1.$

Now we will show that using only half of the Hilbert space, the
positive energy branch, to construct the wavepacket results in a
minimum size of the wavepacket. This minimum size at ${\bf q}=0$
is the Compton wavelength. To start with, we introduce a pair of
projection operators, $\hat{\mathcal{P}}=|u_1\rangle\langle
u_1|+|u_2\rangle\langle u_2|$ and
$\hat{\mathcal{Q}}=|u_3\rangle\langle u_3|+|u_4\rangle\langle
u_4|$. One can see that $\hat{\mathcal{P}}$ projects to positive
energy, $\hat{\mathcal{P}}|w\rangle=|w\rangle$,
$\hat{\mathcal{Q}}$ projects to negative energy, and
$\hat{\mathcal{P}}+\hat{\mathcal{Q}}=1$.

The mean square radius $\Delta_{{\bf r}}$ of the wavepacket in
terms of the projection operators $\hat{\mathcal{P}}$ and
$\hat{\mathcal{Q}}$ is,
\begin{eqnarray}
\Delta_{{\bf r}}^2&\equiv&\langle w| {\bf r}^2|w\rangle-\langle w|
{\bf r}|w\rangle^2\cr &=&\langle w| {\bf
r}(\hat{\mathcal{P}}+\hat{\mathcal{Q}}){\bf r}|w\rangle-\langle w|
{\bf r}|w\rangle^2\cr &=&\langle w| {\bf r}\hat{\mathcal{P}}{\bf
r}|w\rangle-\langle w| {\bf r}|w\rangle^2+\langle w| {\bf
r}\hat{\mathcal{Q}}{\bf r}|w\rangle\cr &=&
\Delta_{\hat{\mathcal{P}}{\bf r}\hat{\mathcal{P}}}^2+\langle w|
{\bf r}\hat{\mathcal{Q}}{\bf r}|w\rangle.
\end{eqnarray}
$\Delta_{\hat{\mathcal{P}}{\bf r}\hat{\mathcal{P}}}$ is the mean
square radius of the projected position operator
$\hat{\mathcal{P}}{\bf r}\hat{\mathcal{P}}$, and is a
positive-definite quantity. The second term is calculated as
follows :
\begin{eqnarray}
&&\langle w|{\bf r}\hat{\mathcal{Q}}{\bf r}|w\rangle\cr
&=&\left(\frac{\lambda_c}{2\epsilon(q_c)}\right)^2
\left|\bar{\mbox{\boldmath$\sigma$}}-\frac{\lambda_c^2}{\epsilon(q_c)
[\epsilon(q_c)+1]}{\bf
q}_c({\bf q}_c\cdot \bar{\mbox{\boldmath$\sigma$}})\right|^2,
\end{eqnarray}
where we have used the relation between the matrix element of
position operator and velocity operator.
$\bar{\mbox{\boldmath$\sigma$}}\equiv\eta_{{\alpha}}^\dagger
\mbox{\boldmath$\sigma$}\eta_{{\alpha}}$ is the spinor-averaged
spin with $\eta_{{\alpha}}= \left(\begin{array}{c}
 \eta_1 \\
  \eta_2
\end{array}\right)$, and $\lambda_c=\frac{\hbar}{mc}$ is the Compton wavelength.

\begin{figure}
\center
\includegraphics[width=3.4in]{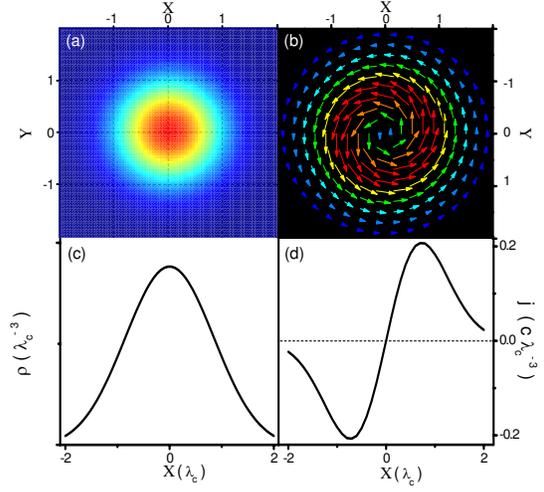}
\caption{(Color online) Distribution of (a) probability density
and (b) probability current density of the wavepacket with a
Gaussian distribution $a({\bf q},t)$. The length scale is in units
of the Compton wavelength $\lambda_c$ and the color bar is from
high density (red) to low density (blue). The profiles of (a) and
(b) along the $x$-axis are plotted in (c) and (d).\label{Fig.1}}
\end{figure}

Thus, we obtain the lower bound of the mean square wavepacket
radius as $\langle w|{\bf r}\hat{\mathcal{Q}}{\bf
r}|w\rangle^{1/2}$. At $q_c=0$, it reduces to half of the Compton
wavelength. We may regard this as the minimum intrinsic radius of
the electron wavepacket. This minimum size is a consequence of
using only half of the Hilbert space in constructing an electron
wavepacket and it is 137 times larger than the classical electron
radius used in Lorentz's argument.\cite{lorentz} Therefore, even
for the tightest possible electron wavepacket, the electron does
not have to rotate faster than the speed of light. To probe the
wavepacket at length scales smaller than the Compton wavelength,
the negative energy branch has to be involved.

In Fig.~1, we plot the probability density, probability current
density of a wavepacket, which are defined as $\rho({\bf
r})=w^\dagger({\bf r})w({\bf r})$ and ${\bf j}({\bf
r})=w^\dagger({\bf r})c\mbox{\boldmath$\alpha$}w({\bf r})$. The
electron wavepacket is spin up (in the $\hat{z}$ direction) and
has a Gaussian distribution $a({\bf q})$ in momentum space with
zero mean momentum ($q_c=0$). A circulating current around the
spin axis is clearly seen in Fig.1b, with maxima at $r=\lambda_c$.
In Fig.~2, the current density of the wavepacket shows a rotating
velocity profile, ${\bf v}({\bf r})={\bf j}({\bf r})/\rho({\bf
r})$, much like that of a rigid sphere (goes linearly with the
radius), except that beyond the edge it gradually saturates to the
speed of light. This implies a rigid core inside the self-rotating
wavepacket. A classical analogy of this is a uniformly charged,
self-rotating sphere, with a diameter of the Compton wavelength,
which is exactly the spinning ball picture of Uhlenbeck and
Goudsmit.\cite{uhlenbeck}

\begin{figure}
\center
\includegraphics[width=3in]{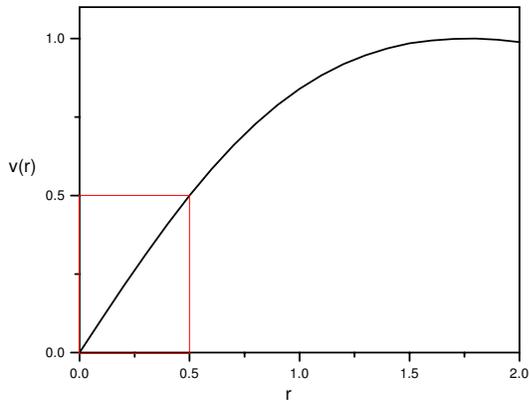}
\caption{Velocity distribution ${\bf v}({\bf r})$ (in units of c)
of a rotating wavepacket. The distance $r$ (in units of the
Compton wavelength $\lambda_c$) measures from the center of
charge. The figure shows the rotating wavepacket has a rigid core
with a diameter equals to the Compton wavelength.}\label{Fig.2}
\end{figure}

\section{The spin magnetic moment}

The current circulating around the spin axis of the wavepacket
would generate a magnetic moment ${\bf M}=\frac{-e}{2}\int d{\bf
r} ({\bf r}-{\bf r}_c)\times {\bf j}({\bf r})$ where ${\bf
r}_c=\langle w|{\bf r}|w\rangle$ is the center of the wavepacket.
With some algebra, one can show that
\begin{eqnarray}
{\bf M}&=&\frac{-e}{2}\langle w|({\bf r}-{\bf r}_c)\times {\bf
v}|w\rangle
\cr&=&\frac{-e}{2}\sum_{{{\alpha\beta}}}\eta^*_{\alpha}(\textbf{q})\mbox{\boldmath${\cal
R}$}_{{{\alpha\beta}}}\times{\bf
v}_{{{\beta\alpha}}}\eta_{\alpha}(\textbf{q}),\label{sp}
\end{eqnarray}
expressed in terms of the matrix element of the velocity operator
${\bf v}_{{\alpha\beta}}=\langle
u_{{\alpha}}|c\mbox{\boldmath$\alpha$}|u_{{\beta}}\rangle$, and
the so-called Berry connection $\mbox{\boldmath${\cal
R}$}_{{\alpha\beta}}=\langle
u_{{\alpha}}|i\frac{\partial}{\partial{\bf
q}}|u_{{\beta}}\rangle$.

After putting in the velocity operator ${\bf
v}=c\mbox{\boldmath$\alpha$}$ in calculation, we obtain
\begin{equation}
{\bf M}=\frac{-e\hbar
}{2m\epsilon^2(q_c)}\left[\bar{\mbox{\boldmath$\sigma$}}+\lambda_c^2\frac{{\bf
q_c \cdot{\bar{\mbox{\boldmath$\sigma$}}}}}{\epsilon(q_c)+1}{\bf
q}_c\right],\label{ma}
\end{equation}
where
$\bar{\mbox{\boldmath$\sigma$}}=\eta_{{\alpha}}^\dagger\mbox{\boldmath$\sigma$}\eta_{{\alpha}}$
is the spinor-average spin. At ${\bf q}_c=0$, it reproduces the
classical result, ${\bf
M}=-\frac{e\hbar}{2m}\bar{\mbox{\boldmath$\sigma$}}=-\mu_B\bar{\mbox{\boldmath$\sigma$}}$,
with the Bohr magneton being $\mu_B=\frac{e\hbar}{2m}$.

In the following, we will show that the magnetic moment induced by
the charge circulation is characterized not by the canonical
angular momentum but by the spin.

The canonical angular momentum operator is defined as ${\bf
L}=m{\bf r}\times{\bf p}=m{\bf r}\times\frac{\hbar}{ i}\nabla$.
Unlike the momentum ${\bf p}$, the canonical angular momentum is
not a conserved quantity, $d{\bf L}/dt\neq0$. It is the total
angular momentum ${\bf J}={\bf L}+{\bf S}$ that is conserved. For
a self-rotating Dirac wavepacket, the canonical angular momentum
is zero (when the momentum operator ${\bf p}$ acts on the
wavefunction $|w\rangle$, it gives $\hbar{\bf q}$, and the matrix
element ${\bf q}_{\alpha\beta}=0$ implies ${\bf L}=\langle w|({\bf
r}-{\bf r}_c)\times{\bf p}|w\rangle=0$).

In Dirac theory, spin is represented as a $4\times 4$ matrix, $
{\mbox{\boldmath$\Sigma$}}=\frac{1}{2}\left(\begin{array}{cc}
 \mbox{\boldmath$\sigma$} & 0 \\
  0 & \mbox{\boldmath$\sigma$}
\end{array}\right).$ We can obtain the average spin by calculating the expectation
value of the spin operator,
\begin{eqnarray}
\bar{\mbox{\boldmath$\Sigma$}}&=&\langle
w|\mbox{\boldmath$\Sigma$}|w\rangle=\sum_{{{\alpha\beta}}}\eta^*_{\alpha}(\textbf{q})\Sigma_{\alpha\beta}\eta_{\beta}(\textbf{q})\cr&=&
\frac{1}{2\epsilon(q_c)}\left[\bar{\mbox{\boldmath$\sigma$}}+\lambda_c^2\frac{{\bf
q_c \cdot{\bar{\mbox{\boldmath$\sigma$}}}}}{\epsilon(q_c)+1}{\bf
q}_c\right],\label{se}
\end{eqnarray}
where $\Sigma_{\alpha\beta}=\langle
u_\alpha|\mbox{\boldmath$\Sigma$}|u_\beta\rangle$. It is
remarkable that the average spin calculated from the abstract spin
operator has the same structure (inside the square bracket of
Eq.(\ref{ma})) as the orbital magnetic moment obtained
semiclassically. We can therefore relate these two quantities by
\begin{equation}
{\bf
M}=-g\frac{e\hbar}{2m\epsilon(q_c)}\bar{\mbox{\boldmath$\Sigma$}},
\end{equation}
where the $g$-factor is 2. Note that the $\epsilon$ in the
denominator can be absorbed in the relativistic mass to form the
relativistic Bohr magneton $\mu_B=e\hbar/2m\epsilon(q_c)$. With
${\bf q}_c=0$, ${\bf M}=-g\mu_B\bar{\mbox{\boldmath$\Sigma$}}$.

The spin therefore can be thought of as coming from the charge
circulation of the electron wavepacket. In fact, the spin is
related to the mechanical angular momentum (the mass circulation
current), ${\bf L}_{mech}=m\langle w|({\bf r}-{\bf r}_c)\times{\bf
v}|w\rangle=2\hbar\bar{\mbox{\boldmath$\Sigma$}}$. The g-factor of
2 is then explained by the fact that the mechanical angular
momentum calculated from the mass circulating current, which is
proportional to the charge circulating current, is twice of the
spin expectation value. In a semiconductor, the g-factor can
deviate from 2 dramatically.\cite{gfactor} The origin of the
anomalous g-factor can be explained as the same way coming from
the self rotation of electron wavepacket.\cite{mcreview}

In the past, there has been a number of attempts to find an
intuitive understanding of the spin magnetic moment within the
framework of the Dirac theory. Huang \cite{huang} suggested that
it can be thought of as the current produced by the zitterbewegung
\cite{zitter}. Ohanian \cite{ohenian} showed that the electron
spin magnetic moment originates from a circulating flow of energy
of the wave field based on an earlier idea of Belinfante
\cite{Bel}.  These ideas are similar in spirits with Uhlenbeck and
Goudsmit's picture of the spin. Here, we see that the rotating
charge model can indeed be re-established explicitly and firmly
within the wavepacket formulation.

The magnetic moment obtained above exists even in the absence of a
magnetic field. We will show that in the existence of an external
magnetic field, the magnetic moment, coming from the self-rotation
of the wavepacket, causes an energy shift in its total energy, the
Zeeman energy.

First we assume the external field is weak and varying on a length
scale much larger than that of the wavepacket. This requirement
allows us to expand the local Hamiltonian around the position of
the charge center ${\bf r}_c$ to the first order of the gradient
correction, $\hat{H}({\bf r}_c,{\bf q}_c,t)=\hat{H}_0({\bf
r}_c,{\bf q}_c,t)+({\bf r}-{\bf r}_c)\cdot(\partial
\hat{H}/\partial{\bf r}_c).$ For a uniform magnetic field ${\bf
B}=\nabla\times{\bf A}({\bf r}_c,t)$, with a symmetric vector
potential ${\bf A}({\bf r},t)=\frac{1}{2}{\bf B}\times{\bf r}$, we
have $\hat{H}_0({\bf r}_c,{\bf
k}_c,t)=c\mbox{\boldmath$\alpha$}\cdot\hbar{\bf k}_c+\beta mc^2$,
where ${\bf k}_c={\bf q}_c+\frac{e}{\hbar}{\bf A}({\bf r}_c,t)$ is
the kinetic momentum. The energy correction due to the external
field is given by
\begin{eqnarray}
\delta E&=&\left\langle w\left|{({\bf r}-\bf r}_c)\cdot\partial
\hat{H}/\partial{\bf r}_c\right|w\right\rangle\cr &=&\left\langle
w|\Sigma_{i,j}(r_i-r_{c,i})e\partial A_j/\partial
r_i|w\right\rangle=-{\bf M}\cdot {\bf B},
\end{eqnarray}
We therefore observe that the Zeeman energy comes from the energy
gradient correction and is associated with the ${\bf M}$ defined
in Eq.(\ref{sp}).

When both the electric field $\bf{E}$ and magnetic field $\bf{B}$
are present, the total energy of a wavepacket is
\begin{equation}
E({\bf r}_c,{\bf k}_c)=\langle w|\hat{H}|w\rangle=E_0({\bf
k}_c)-e\phi({\bf r}_c)-{\bf M}\cdot {\bf B},\label{E}
\end{equation}
where $E_0({\bf k}_c)$ is given by Eq.~(\ref{dis}) with
$q_c\rightarrow k_c$, and $\phi({\bf r}_c)$ is the scalar
potential of the electric field.

\section{The dynamics of the wavepacket}

Surprisingly, there are no spin-orbit coupling in the wavepacket
energy (Eq.~(\ref{E})), which is expected to appear at the first
order of the electric field. One can quantize the semiclassical
Dirac electron and show that the spin-orbit coupling is related to
the non-canonical wavepacket
dynamics.\cite{mcreview,bliokh,berard,xiao} The effective
Lagrangian of a wavepacket is \cite{culcer},
\begin{equation}
\mbox{\boldmath${\cal L}$}=i\hbar\eta^\dagger\frac{\partial
\eta}{\partial t}+\hbar\dot\textbf{k}_c\cdot\mbox{\boldmath${\cal
R}$}+\hbar \textbf{k}_c\cdot{\dot{\textbf
r}_c}-\frac{e}{c}\textbf{A}\cdot{\dot{\textbf{r}}_c}-E(\textbf{r}_c,\textbf{k}_c).
\label{Leff}
\end{equation}
For a Dirac electron, the Berry connection is
$\mbox{\boldmath${\cal
R}$}=\frac{\lambda_c^2}{2\epsilon(\epsilon+1)}\textbf{
k}_c\times{{\mbox{\boldmath$\sigma$}}}.\label{re2}$

From the Lagrangian, one can derive the equations of motion for
the centers of charge position and momentum, correct to linear
order in fields,
\begin{eqnarray}
\hbar\dot\textbf{k}_c&=&-e\textbf{E}-\frac{e}{c}\frac{\hbar\textbf{
k}_c}{\epsilon m}\times\textbf{B},\label{k}\\
\dot\textbf{r}_c&=&\frac{\hbar\textbf{k}_c}{\epsilon
m}+\frac{e}{\hbar}\left(\textbf{E}\times\textbf{F} +\textbf{
B}\cdot\textbf{F} \frac{{\hbar\textbf{k}_c}}{\epsilon
mc}\right),\label{r}
\end{eqnarray}
where $\textbf{F}=\langle\mbox{\boldmath${\cal
F}$}\rangle=\eta_\alpha^\dagger\mbox{\boldmath${\cal
F}$}\eta_\alpha$, $\mbox{\boldmath${\cal F}$}
=-\frac{\lambda_c^2}{2\epsilon^3}\left(\mbox{\boldmath$\sigma$}+\lambda_c^2\frac{\textbf{
k}_c \cdot\mbox{\boldmath$\sigma$}}{\epsilon+1}\textbf{
k}_c\right)$ is called the Berry curvature.

The equation for spin precession is given by
\begin{equation}
\dot{\bar{\mbox{\boldmath$\sigma$}}}=(\lambda_c/\epsilon)(e/\hbar)\left[\textbf{B}
+(\lambda_c/(\epsilon+1))\textbf{E}\times\textbf{k}_c
\right]\times\bar{\mbox{\boldmath$\sigma$}},
\end{equation}
which agrees with the Bargmann-Michel-Telegdi equation\cite{bmt}.

\begin{figure}
\center
\includegraphics[width=3in]{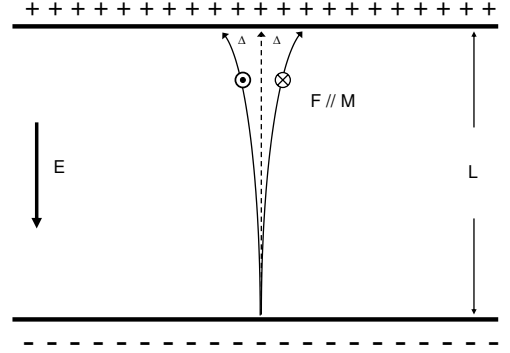}
\caption{Electrons are accelerated by a static electric field in a
parallel plate capacitor. Magnetic moment $\textbf{M}$ points into
(out of) the page, giving rise to an anomalous velocity to the
right (left). The trajectory shift is
$\Delta=\lambda_c\sqrt{eV_0/2mc^2}$ with a potential difference
$V_0=EL$.}\label{Fig.3}
\end{figure}

When only the electric field exists ($\textbf{B}=0$ in
Eq.~(\ref{r})), we find that the wavepacket has an anomalous
velocity in the direction of ${\bf E}\times{\bf F}$, and since
$\mbox{\boldmath${\cal F}$}\propto \mbox{\boldmath$\sigma$}$ at
low velocity, spin-up and spin-down electrons would have opposite
transverse velocities (see Fig.~\ref{Fig.3}).

Notice that the $\textbf{r}_c$ and $\textbf{k}_c$ in Eq.~(14) are
not a canonical pair, due to the presence of the gauge potentials
$\mbox{\boldmath${\cal R}$}$ and $\textbf{A}$. Their connections
with canonical variables ${\bf r}$ and ${\bf p }$ are given by
(valid in weak fields), \cite{mcreview}
\begin{eqnarray}
\textbf{r}_c&=&\textbf{r}+\mbox{\boldmath${\cal
R}$}(\mbox{\boldmath$\pi$})+\mbox{\boldmath${\cal G}$}(\textbf{k}_c)(\mbox{\boldmath$\pi$}),\nonumber\\
\hbar\textbf{k}_c&=&\mbox{\boldmath$\pi$}+\frac{e}{c}\textbf{
B}\times\mbox{\boldmath${\cal
R}$}(\mbox{\boldmath$\pi$}),\label{extra}
\end{eqnarray}
where
$\mbox{\boldmath$\pi$}=\textbf{p}+\frac{e}{c}\textbf{A}(\textbf{
r})$, and $\mbox{\boldmath${\cal G}$}_\alpha\equiv 1/2(\partial
\mbox{\boldmath${\cal R}$}/\partial
k^\alpha)\cdot(\mbox{\boldmath${\cal R}$}\times\textbf{B})$. This
is analogous to the Peierls substitution for the momentum.

Now we can {\it re-quantize} the semiclassical Dirac energy
Eq.~(\ref{E}) and obtain the relativistic Pauli Hamiltonian for
all orders of velocity\cite{silenko},
\begin{equation}
{H}(\textbf{r},\textbf{p})=\epsilon({\pi})
mc^2-e\phi(\textbf{r})+\frac{\mu_B}{\epsilon}\mbox{\boldmath$\sigma$}\cdot
\left[\frac{\textbf{E}\times\mbox{\boldmath$\pi$}}{(\epsilon+1)mc}+\textbf{B}\right]
\label{pa}
\end{equation}
This alternative approach is intuitive when compared to formal
procedures of block-diagonalization, such as the Foldy-Wouthuysen
transformation.\cite{foldy}

In Eq.~(\ref{pa}), the third term is the spin-orbit coupling which
emerges from the first-order gradient expansion of the scalar
potential,
$\partial\phi/\partial\textbf{r}\cdot\mbox{\boldmath${\cal R}$}$.
In the literature, $-e\mbox{\boldmath${\cal R}$}$ has often been
called an electric dipole which couples to the electric field to
give rise to the spin-orbit energy\cite{mathur}. For electrons in
narrow gap semiconductors, the spin-orbit coupling is called a
Yafet term\cite{yafet}. This is unfortunately artificial, because
its existence depends on the unphysical position $\textbf{r}$
which depends on the choice of the SU(2) gauge, instead of the
true position $\textbf{r}_c$. The equations of motion based on the
Pauli Hamiltonian is consistent with the Dirac theory if and only
if one recognizes this fact.

\section{Spin Hall Effect and Spin Nernst Effect}

The presence of the Berry curvature gives the Dirac electron a
tiny but nonzero anomalous velocity in the vacuum. Similar to the
electron in semiconductor, such a Berry curvature would lead to
the spin Hall effect and the spin Nernst effect. The discussion
below relies on the formulation developed previously for the Hall
effect and the Nernst effect for spinless electrons.\cite{xiao}
But their results are strictly applicable to the present case as
long as the electron spin is conserved.

For spinless electrons, the Hall conductivity is given by
\begin{equation}
\sigma_{xy}=-\frac{e^2}{\hbar}\int \frac{d^3 k}{(2\pi)^3}f({\bf
k})\Omega_z({\bf k}),\label{hall}
\end{equation}
where $f({\bf k})$ is the Fermi distribution function in
equilibrium, $\Omega_z({\bf k})$ is the Abelian Berry curvature.
The Nernst current perpendicular to the temperature gradient is
given by $ j_x=\alpha_{xy}(-\nabla_y T)$, and the Nernst
coefficient $\alpha_{xy}$ is related to the Hall conductivity
$\sigma_{xy}$ via
\begin{equation}
\alpha_{xy}=\frac{1}{e}\int dE\left(-\frac{\partial f}{\partial E
}\right)\sigma_{xy}(E)\frac{E-\mu}{T},\label{nernst}
\end{equation}
where $\sigma_{xy}(E)$ is the Hall conductivity from all of the
states below energy $E$ and $\mu$ is the chemical potential.

For electrons with spins, we need to replace the Abelian Berry
curvature $\Omega_z$ in Eq.~(\ref{hall}) by the non-Abelian one
averaged over spin, $\langle {\cal F}_z\rangle$. For a Dirac
electron, in the limit of $\hbar k<<mc$, we have ${\cal
F}_z=-(\lambda_c^2/2)\sigma_z$. For a Dirac electron gas that is
not spin-polarized, the spin-averaged $\langle {\cal F}_z\rangle$
is zero, even though ${\cal F}_z$ itself is non-zero. As a result,
one expects neither charge Hall effect nor Nernst effect.

If the electron gas is spin polarized, then $\langle {\cal
F}_z\rangle$ is not zero and one has the anomalous Hall effect
(see Eq.~(\ref{hall})). At the mean time, according to
Eq.~(\ref{nernst}), there is an anomalous Nernst effect. Again in
the small momentum limit, we have
\begin{equation}
\sigma_{xy}\simeq
\frac{e^2}{\hbar}\frac{\lambda_c^2}{2}\frac{n}{2}\langle\sigma_z\rangle,
\end{equation}
where $n$ is the electron density.

At low temperature (compared to the Fermi temperature), the Nernst
coefficient and the Hall coefficient are related by the Mott
relation (which can be derived from Eq.~(\ref{nernst})),
\begin{equation}
\alpha_{xy}\simeq
\frac{\pi^2}{3}\frac{k_B^2T}{e}\frac{d\sigma_{xy}}{d\mu}.
\end{equation}
Therefore, $\alpha_{xy}$ is proportional to the density of states
at the Fermi level, $dn/d\mu$.

Since the Berry curvature ${\cal F}_z\propto\sigma_z$ at low
velocity, spin-up and spin-down electrons would move to opposite
transverse directions. Therefore, even if the electron gas is not
spin polarized, there can still be a spin Hall effect (see Fig.3).
This is analogous to the emergence of the spin Hall effect in bulk
(non-magnetic) semiconductors.\cite{murakami}

The spin Hall conductivity is given as (valid when electron spin
is conserved),
\begin{equation}
\sigma^z_{xy}=\frac{e}{2}\int \frac{d^3 k}{(2\pi)^3}f({\bf
k})\langle\sigma_z {\cal F}_z\rangle,
\end{equation}
which is approximately equal to $(e/2)(\lambda_c^2/2)(n/2)$.
Similarly, the spin Nernst coefficient is given by
\begin{equation}
\alpha^z_{xy}\simeq
\frac{\pi^2}{3}\frac{k_B^2T}{e}\frac{d\sigma^z_{xy}}{d\mu}.
\end{equation}
For Fermi gas at low temperature, we have
$\sigma^z_{xy}=\frac{ek_F^3\lambda_c^2}{24\pi^2}$ and
$\alpha^z_{xy}\simeq\frac{k_B^2k_F^2\lambda_c^2}{24}T$. For
electrons in semiconductor, the Berry curvature has the same
structure as in vacuum but with different coefficient, i.e. ${\cal
F}_z=\frac{2V^2}{3}[\frac{1}{E_g^2}-\frac{1}{(E_g+\Delta)^2}]\sigma_z$,
therefore the effect can be enlarged by a factor
$\frac{4V^2}{3\lambda_c^2}[\frac{1}{E_g^2}-\frac{1}{(E_g+\Delta)^2}]$,
where $E_g$ is the energy gap between conduction band and top
valence band, and $\Delta$ is energy separation between the
split-off band and top valence band. For example, in GaAs with
$E_g=1.424$eV, $\Delta=0.34$eV, $E_p=2m_eV^2/\hbar^2=22.7$eV, the
effect is enlarged by $1.3\times 10^6$ times. Similar to the
anomalous Nernst effect, such a contribution is ultimately
originated from the Berry phase correction to (now spin-dependent)
orbital magnetization.\cite{xiao}

\section{Conclusion}

We have shown that a self-rotation picture of the wavepacket
explains the origin of the electron spin by regarding the
non-relativistic electron as a wavepacket at the bottom of the
positive energy branch of the Dirac theory. The minimum size of
the wavepacket equals to the Compton wavelength. The magnetic
moment generated from the circulating charge current gives the
Bohr magneton in non-relativistic limit, and is responsible for
the Zeeman energy under the external fields. The g-factor of 2
comes from the fact that the mechanical angular momentum from the
mass circulating current is twice of the spin expectation value.
The spin-orbit coupling emerges from the first-order gradient
expansion of the scalar potential and is related to the Berry
connection. Finally, the Berry curvature plays an important role
in both the spin Hall effect and the spin Nernst effect. Although
the predictions of our semiclassical theory can be calculated from
the microscopic Dirac theory, it provides not only an intuitive
conceptual view but also a quantitatively accurate theoretical
framework. The method can be directly transplanted to Bloch
electrons in crystals, making predictions on various
thermodynamics as well as transport phenomena, such as spin Nernst
effect discussed specifically in this paper.

The authors would like to thank E. I. Rashba, M. Stone, L. Balent,
D. Culcer, and D. Xiao for many helpful discussions.

\end{document}